\magnification \magstep1
\raggedbottom
\openup 2\jot
\voffset6truemm
\def\cstok#1{\leavevmode\thinspace\hbox{\vrule\vtop{\vbox{\hrule\kern1pt
\hbox{\vphantom{\tt/}\thinspace{\tt#1}\thinspace}}
\kern1pt\hrule}\vrule}\thinspace}
\centerline {\bf BOUNDARY OPERATORS IN QUANTUM FIELD THEORY}
\vskip 1cm
\leftline {Giampiero Esposito}
\vskip 0.3cm
\noindent
{Istituto Nazionale di Fisica Nucleare, Sezione di Napoli,
Complesso Universitario di Monte S. Angelo, Via Cintia, Edificio N',
80126 Napoli, Italy}
\vskip 0.3cm
\noindent
{Universit\`a di Napoli Federico II, Dipartimento di Scienze
Fisiche, Complesso Universitario di Monte S. Angelo, Via Cintia,
Edificio N', 80126 Napoli, Italy}
\vskip 1cm
\noindent
The fundamental laws of physics can be derived 
from the requirement of invariance under suitable classes of
transformations on the one hand, and from the need for a well-posed
mathematical theory on the other hand. As a part of this programme,
the present paper shows under which conditions the introduction of
pseudo-differential boundary operators in one-loop Euclidean 
quantum gravity is compatible both with their invariance under 
infinitesimal diffeomorphisms and with the requirement of a strongly
elliptic theory. Suitable assumptions on the kernel of the boundary
operator make it therefore possible to overcome problems resulting
from the choice of purely local boundary conditions.
\vskip 1cm
\leftline {PACS numbers: 03.70.+k, 04.70.Gw}
\vskip 100cm
\leftline {\bf 1. Introduction}
\vskip 0.3cm
The aim of theoretical physics is to
provide a clear conceptual framework for the wide variety of natural 
phenomena, so that not only are we able to make accurate predictions
to be checked against observations, but the underlying mathematical
structures of the world we live in can also become sufficiently 
well understood by the scientific community. What are therefore the
key elements of a mathematical description of the physical world?
Can we derive all basic equations of theoretical physics from a few
symmetry principles? What do they tell us about the origin and
evolution of the universe? Why is gravitation so peculiar with
respect to all other fundamental interactions?

The above questions have received careful consideration over the last
decades, and have led, in particular, to several approaches to a
theory aiming at achieving a synthesis of quantum physics on the one
hand, and general relativity on the other hand. This remains, possibly, 
the most important task of theoretical physics. 
The need for a quantum theory of gravity is already clear
from singularity theorems in classical cosmology. Such theorems [1] 
prove that the Einstein theory of general relativity leads to the
occurrence of space-time singularities in a generic way. 
At first sight one might be tempted to conclude
that a breakdown of all physical laws occurred in the past,
or that general relativity is severely incomplete, being unable to
predict what came out of a singularity. It has been therefore 
suggested that all these pathological features result from the attempt
of using the Einstein theory well beyond its limit of validity, i.e.
at energy scales where the fundamental theory is definitely more 
involved. General relativity might be therefore viewed as a low-energy
limit of a richer theory, which achieves the synthesis of both the
basic principles of modern physics and the fundamental interactions
in the form presently known [2]. 

Within the framework just outlined it remains however true that the
various approaches to quantum gravity developed so far suffer from
mathematical inconsistencies, or incompleteness in their ability
of accounting for some basic features of the laws of nature. From the
point of view of general principles, the space-time approach to 
quantum mechanics and quantum field theory [3--5], and its application
to the quantization of gravitational interactions, remains indeed of
fundamental importance [6,7]. When one tries to implement the Feynman
``sum over histories'' one discovers that, already at the level of
non-relativistic quantum mechanics, a well defined mathematical
formulation is only obtained upon considering a heat-equation
problem. The measure occurring in the Feynman representation of the
Green kernel is then meaningful, and the propagation amplitude of
quantum mechanics in flat Minkowski space-time is obtained by
analytic continuation. This is a clear indication that 
quantum-mechanical problems via path integrals are well understood 
only if the heat-equation counterpart is mathematically well posed.
In quantum field theory one then deals with the Euclidean approach,
and its application to quantum gravity relies heavily on the theory
of elliptic operators on Riemannian manifolds [8]. To obtain a complete
picture one has then to specify the boundary conditions of the 
theory, i.e. the class of Riemannian geometries with their topologies
involved in the sum, and the form of boundary data assigned on
the bounding surfaces.

In particular, recent work [9] has shown
that the only set of local boundary conditions on metric
perturbations which are completely invariant under infinitesimal
diffeomorphisms is incompatible with the request of a good elliptic
theory. More precisely, while the resulting operator on metric
perturbations can be made of Laplace type and elliptic in the
interior of the Riemannian manifold under consideration, the property
of strong ellipticity is violated. This is a precise 
mathematical expression of the request that a unique smooth solution
of the boundary-value problem should exist which vanishes at infinite
geodesic distance from the boundary. This opens deep interpretive
issues, since only for gravity does the request of complete gauge
invariance of the boundary conditions turn out to be incompatible
with a good elliptic theory [9]. It is then impossible to make sense
even just of the one-loop semiclassical approximation, because the
functional trace of the heat operator is found to diverge [9].

We have been therefore led to consider non-local
boundary conditions for the quantized gravitational field at one-loop
level [10,11]. On the one hand, such a 
scheme already arises in simpler
problems, i.e. the quantum theory of a free particle subject to 
non-local boundary data on a circle [12]. One then finds two families of
eigenfunctions of the Hamiltonian: {\it surface states} which decrease
exponentially as one moves away from the boundary, and 
{\it bulk states} which remain instead smooth and non-vanishing. 
The generalization to an Abelian gauge theory such as Maxwell theory
can fulfill non-locality, ellipticity and complete gauge invariance
of boundary conditions providing one learns to work with  
pseudo-differential operators in one-loop quantum theory [13]. On the
other hand, in the application to quantum gravity, since the boundary
operator acquires new kernels responsible for the pseudo-differential
nature of the boundary-value problem, one might hope to be able to
recover a good elliptic theory under a wider variety of conditions. 

This is precisely the aim of the present paper. After a survey of
operators of Laplace type and of the associated boundary operators
in section 2, section 3 introduces integro-differential boundary
operators in Euclidean quantum gravity. Strong ellipticity of
differential and pseudo-differential boundary-value problems is
then defined in detail in section 4, and the application to 
Euclidean quantum gravity is studied in section 5. Further examples,
of simpler nature, are given in section 6, and concluding remarks
are presented in section 7.
\vskip 0.3cm
\leftline {\bf 2. Operators of Laplace type and their boundary 
operators} 
\vskip 0.3cm
In the Euclidean approach to quantum field theory and quantum gravity
one studies differentiable manifolds endowed with 
positive-definite metrics $g$, so that space-time is actually
replaced by an $m$-dimensional Riemannian space $(M,g)$. An operator
${\cal P}$ of Laplace type, which acts on gauge fields, maps smooth 
sections of a vector bundle $V$ over $M$ into smooth sections of
the same bundle, i.e.
$$
{\cal P}: C^{\infty}(V,M) \rightarrow C^{\infty}(V,M),
$$
and reads
$$
{\cal P}=-g^{ab}\nabla_{a}\nabla_{b}-E,
\eqno (2.1)
$$
where $g^{ab}$ is the contravariant form of the Riemannian metric
for $M$, $\nabla$ is the connection on $V$, and $E$ is an 
endomorphism. In Ref. [9] a thorough investigation of boundary 
operators for elliptic operators of the form (2.1) has been
performed. The key elements we need to recall are as follows.

If the manifold $M$ has a smooth non-empty boundary $\partial M$,
two vector bundles over $\partial M$, hereafter denoted by $W$ and $W'$,
yield a complete description of the problem. The boundary operator
$B$ maps smooth sections of $W$ into smooth sections of $W'$:
$$
B: C^{\infty}(W,\partial M) \rightarrow 
C^{\infty}(W',\partial M).
$$
For mixed boundary conditions, the 
operator $B$ frequently reads [9,14,15]
$$
B \equiv \pmatrix{\Pi & 0 \cr \Lambda & I-\Pi \cr},
\eqno (2.2)
$$
where $\Pi$ and $I-\Pi$ are complementary projectors, and $\Lambda$
is a first-order tangential differential operator
$$
\Lambda \equiv
(I-\Pi) \left[{1\over 2}\Bigr(\Gamma^{i}{\widehat \nabla}_{i}
+{\widehat \nabla}_{i}\Gamma^{i}\Bigr)+S \right](I-\Pi).
\eqno (2.3)
$$
With our notation, $\Gamma^{i}$ are endomorphism-valued vector fields on
the boundary, $\widehat \nabla$ is the induced connection on 
$\partial M$, and $S$ is an endomorphism on $\partial M$. By 
virtue of (2.3) one has
$$
\Pi \Lambda=\Lambda \Pi=0,
\eqno (2.4)
$$
and hence $B$ is a projector, in that $B^{2}=B$. The boundary-value
problem is meant to be the pair $({\cal P},B)$, 
where ${\cal P}$ is the operator
(2.1) and $B$ is given in (2.2). The corresponding mixed boundary
conditions read
$$
\pmatrix{\Pi & 0 \cr \Lambda & I-\Pi \cr}
\pmatrix{[\varphi]_{\partial M} \cr
[\varphi_{;N}]_{\partial M} \cr}=0,
\eqno (2.5)
$$
where ${ }_{;N}$ denotes covariant differentiation along the direction
normal to the boundary, i.e. $N^{a}\nabla_{a}$. Moreover, the
boundary operator (2.2) may be expressed in the form
$$
B=PL,
\eqno (2.6)
$$
where $P$ is the map
$$
P:C^{\infty}(W,\partial M) \rightarrow C^{\infty}(W',\partial M)
$$
given by
$$
P \equiv \pmatrix{\Pi & 0 \cr 0 & I-\Pi \cr},
\eqno (2.7)
$$
and $L$ is a map
$$
L: C^{\infty}(W,\partial M) \rightarrow C^{\infty}(W,\partial M)
$$
expressed in matrix form as
$$
L \equiv \pmatrix{I & 0 \cr \Lambda & I \cr}.
\eqno (2.8)
$$
Interestingly, the operator $P$ is itself a projector: $P^{2}=P$,
whereas $L$ is non-singular, with inverse
$$
L^{-1}=\pmatrix{I & 0 \cr -\Lambda^{-1} & I \cr}.
\eqno (2.9)
$$
The ``column vector'' used in Eq. (2.5), i.e.
$$
\psi(\varphi) \equiv \pmatrix{[\varphi]_{\partial M} \cr
[\varphi_{;N}]_{\partial M} \cr},
\eqno (2.10)
$$
is a section of the bundle $W$ of boundary data, whereas the
auxiliary vector bundle $W'$ has sections given by (see the main
diagonal of $B$ in (2.2))
$$
\psi'(\varphi) \equiv \pmatrix{\Pi[\varphi]_{\partial M} \cr
(I-\Pi)[\varphi_{;N}]_{\partial M} \cr}.
\eqno (2.11)
$$
At this stage, a naturally occurring question is 
under which conditions a projector $P$
gives rise to a projector $B$ such that $B=PL$ as in Eq. (2.6).
To obtain equations in a form as general as possible we replace 
$P$ defined in (2.7) by the $2 \times 2$ matrix
$$
P \equiv \pmatrix{\alpha & \beta \cr \gamma & \delta \cr},
\eqno (2.12)
$$
where $\alpha,\beta,\gamma,\delta$ are, for the time being, some
unknown operators to be determined by imposing suitable 
restrictions (see below). The projector condition $P^{2}=P$
yields therefore four operator equations, i.e.
$$
\alpha^{2}+\beta \gamma=\alpha,
\eqno (2.13)
$$
$$
\alpha \beta + \beta \delta =\beta,
\eqno (2.14)
$$
$$
\gamma \alpha + \delta \gamma=\gamma,
\eqno (2.15)
$$
$$
\gamma \beta + \delta^{2}=\delta.
\eqno (2.16)
$$
A particular solution of Eqs. (2.13)--(2.16) is given by the case
in which
$$
\beta=\gamma=0,
\eqno (2.17)
$$
$$
\alpha^{2}=\alpha,
\eqno (2.18)
$$
$$
\delta^{2}=\delta,
\eqno (2.19)
$$
$$
\alpha+\delta=I.
\eqno (2.20)
$$
This yields the operator $P$ in the form (2.7) appropriate for the
Grubb--Gilkey--Smith boundary-value problem [14,15]. If the
conditions (2.17)--(2.20) are not fulfilled, one gets
instead from Eqs. (2.13)--(2.16) the equations
$$
\alpha(\alpha-I)=-\beta \gamma,
\eqno (2.21)
$$
$$
\alpha=I-\beta \delta \beta^{-1},
\eqno (2.22)
$$
$$
\alpha=I-\gamma^{-1}\delta \gamma,
\eqno (2.23)
$$
$$
\delta(\delta-I)=-\gamma \beta,
\eqno (2.24)
$$
provided that $\beta$ and $\gamma$ can be inverted.
\vskip 0.3cm
\leftline {\bf 3. Euclidean quantum gravity}
\vskip 0.3cm
In Euclidean quantum gravity, mixed boundary conditions on metric
perturbations $h_{cd}$ occur naturally if one requires their
complete invariance under infinitesimal diffeomorphisms, as is
proved in detail in Ref. [9]. On denoting by $N^{a}$ the inward-pointing
unit normal to the boundary, by
$$
q_{\; b}^{a} \equiv \delta_{\; b}^{a}-N^{a}N_{b}
\eqno (3.1)
$$
the projector of tensor fields 
onto $\partial M$, with associated projection operator
$$
\Pi_{ab}^{\; \; \; cd} \equiv q_{\; (a}^{c} \; q_{\; b)}^{d},
\eqno (3.2)
$$
the gauge-invariant boundary conditions for one-loop quantum gravity
read [9]
$$
\Bigr[\Pi_{ab}^{\; \; \; cd}h_{cd}\Bigr]_{\partial M}=0,
\eqno (3.3)
$$
$$
\Bigr[\Phi_{a}(h)\Bigr]_{\partial M}=0,
\eqno (3.4)
$$
where $\Phi_{a}$ is the gauge-averaging functional necessary to
obtain an invertible operator $P_{ab}^{\; \; \; cd}$ on metric
perturbations. When $P_{ab}^{\; \; \; cd}$ is chosen to be of
Laplace type, $\Phi_{a}$ reduces to the familiar de Donder term
$$
\Phi_{a}(h)=\nabla^{b}\Bigr(h_{ab}-{1\over 2}g_{ab}g^{cd}
h_{cd}\Bigr)=E_{a}^{\; bcd}\nabla_{b}h_{cd},
\eqno (3.5)
$$
where $E^{abcd}$ is the DeWitt supermetric on the vector bundle
of symmetric rank-two tensor fields over $M$
($g$ being the metric on $M$):
$$
E^{abcd} \equiv {1\over 2}\Bigr(g^{ac}g^{bd}+g^{ad}g^{bc}
-g^{ab}g^{cd}\Bigr).
\eqno (3.6)
$$
The boundary conditions (3.3) and (3.4) can then be cast in the
Grubb--Gilkey--Smith form (2.5), where $\Lambda$ is the first-order
operator on the boundary defined in Eq. (2.3). However, the
work in Ref. [9] has shown that an operator of Laplace type on
metric perturbations is then incompatible with the requirement 
of strong ellipticity of the boundary-value problem (see section 4),
because the operator $\Lambda$ contains tangential derivatives
of metric perturbations.

To take care of this serious drawback, the work in Refs. [10,11] has 
proposed to consider in the boundary condition (3.4) a 
gauge-averaging functional given by the de Donder term (3.5) plus
an integro-differential operator on metric perturbations, i.e.
$$
\Phi_{a}(h) \equiv E_{a}^{\; bcd}\nabla_{b}h_{cd}
+\int_{M}\zeta_{a}^{\; cd}(x,x')h_{cd}(x')dV'.
\eqno (3.7)
$$
We now begin by remarking that the resulting boundary conditions 
can be cast in the form
$$
\pmatrix{\Pi & 0 \cr \Lambda+{\widetilde \Lambda} & I-\Pi \cr}
\pmatrix{[\varphi]_{\partial M} \cr [\varphi_{;N}]_{\partial M} \cr}
=0,
\eqno (3.8)
$$
where $\widetilde \Lambda$ reflects the occurrence of the integral
over $M$ in Eq. (3.7). It is convenient to work first in a
general way and then consider the form taken by these operators 
in the gravitational case. On requiring that the resulting 
boundary operator
$$
{\cal B} = \pmatrix{{\cal B}_{11} & {\cal B}_{12} \cr
{\cal B}_{21} & {\cal B}_{22} \cr} 
\equiv \pmatrix{\Pi & 0 \cr \Lambda+{\widetilde \Lambda}
& I-\Pi \cr}
\eqno (3.9)
$$
should remain a projector: ${\cal B}^{2}={\cal B}$, we find
the condition
$$
(\Lambda+{\widetilde \Lambda})\Pi
-\Pi(\Lambda+{\widetilde \Lambda})=0,
\eqno (3.10)
$$
which reduces to
$$
\Pi {\widetilde \Lambda}={\widetilde \Lambda}\Pi,
\eqno (3.11)
$$
by virtue of (2.4).

In Euclidean quantum gravity at one-loop level, Eq. (3.11) 
leads to
$$ 
\Pi_{a \; \; c}^{\; b \; \; r}(x) \int_{M}
\zeta_{b}^{\; cq}(x,x')h_{qr}(x')dV' 
=\int_{M}\zeta_{a}^{\; cd}(x,x')\Pi_{cd}^{\; \; \; qr}(x')
h_{qr}(x')dV',
\eqno (3.12)
$$   
which can be re-expressed in the form
$$
\int_{M}\left[\Pi_{a \; \; c}^{\; b \; \; r}(x)
\zeta_{b}^{\; cq}(x,x')-\zeta_{a}^{\; cd}(x,x')
\Pi_{cd}^{\; \; \; qr}(x')\right]h_{qr}(x')dV'=0.
\eqno (3.13)
$$
Since this should hold for all $h_{qr}(x')$, it eventually leads
to the vanishing of the term in square brackets in the integrand.
The notation $\zeta_{b}^{\; cq}(x,x')$ is indeed rather
awkward, because there is an even number of arguments, i.e.
$x$ and $x'$, with an odd number of indices. Hereafter, we
therefore assume that a vector field $T$ and kernel
$\widetilde \zeta$ exist such that
$$
\zeta_{b}^{\; cq}(x,x') \equiv T^{p}(x)
{\widetilde \zeta}_{bp}^{\; \; \; cq}(x,x')
\equiv T^{p}{\widetilde \zeta}_{bp}^{\; \; \; c'q'}.
\eqno (3.14)
$$
The projector condition (3.11) is therefore satisfied if and 
only if
$$
T^{p}(x)\left[\Pi_{a \; \; c}^{\; b \; \; r}(x)
{\widetilde \zeta}_{bp}^{\; \; \; cq}(x,x')
-{\widetilde \zeta}_{ap}^{\; \; \; cd}(x,x')
\Pi_{cd}^{\; \; \; qr}(x')\right]=0.
\eqno (3.15)
$$
\vskip 0.3cm
\leftline {\bf 4. Strong ellipticity}
\vskip 0.3cm
We are here concerned with the issue of ellipticity of the
boundary-value problem of section 3. For this purpose, we begin
by recalling what is known about ellipticity of the Laplacian 
(hereafter $P$) on a Riemannian manifold with smooth boundary.
This concept is studied in terms of the leading symbol of $P$.
It is indeed well known that the Fourier transform makes it
possible to associate to a differential operator of order $k$
a polynomial of degree $k$, called the characteristic polynomial
or symbol. The leading symbol, $\sigma_{L}$, picks out the
highest order part of this polynomial. For the Laplacian,
it reads
$$
\sigma_{L}(P;x,\xi)=|\xi|^{2}I=g^{\mu \nu}\xi_{\mu}\xi_{\nu}I.
\eqno (4.1)
$$
With a standard notation, $(x,\xi)$ are local coordinates
for $T^{*}(M)$, the cotangent bundle of $M$. The leading symbol
of $P$ is trivially elliptic in the interior of $M$, since the
right-hand side of (4.1) is positive-definite, and one has
$$
{\rm det}\Bigr[\sigma_{L}(P;x,\xi)-\lambda \Bigr]
=(|\xi|^{2}-\lambda)^{{\rm dim} \; V} \not = 0,
\eqno (4.2)
$$
for all $\lambda \in {\cal C}-{\bf R}_{+}$. In the presence of
a boundary, however, one needs a more careful definition of
ellipticity. First, for a manifold $M$ of dimension $m$, the
$m$ coordinates $x$ are split into $m-1$ local coordinates on
$\partial M$, hereafter denoted by $\left \{ {\hat x}^{k} 
\right \}$, and $r$, the geodesic distance to the boundary. 
Moreover, the $m$ coordinates $\xi_{\mu}$ are split into
$m-1$ coordinates $\left \{ \zeta_{j} \right \}$ (with $\zeta$
being a cotangent vector on the boundary), jointly with a real
parameter $\omega \in T^{*}({\bf R})$. At a deeper level, all this
reflects the split
$$
T^{*}(M)=T^{*}({\partial M})\oplus T^{*}({\bf R})
\eqno (4.3)
$$
in a neighbourhood of the boundary [8,9].

The ellipticity we are interested in requires now that $\sigma_{L}$
should be elliptic in the interior of $M$, as specified before, and
that strong ellipticity should hold. This means that a unique
solution exists of the differential equation obtained from
the leading symbol:
$$
\left[\sigma_{L}\left(P; \left \{ {\hat x}^{k} \right \}, r=0,
\left \{ \zeta_{j} \right \}, \omega \rightarrow -i
{\partial \over \partial r} \right)-\lambda \right]
\varphi(r,{\hat x},\zeta;\lambda)=0,
\eqno (4.4)
$$
subject to the boundary conditions
$$
\sigma_{g}(B)\left( \left \{ {\hat x}^{k} \right \},
\left \{ \zeta_{j} \right \} \right) \psi(\varphi)
=\psi'(\varphi)
\eqno (4.5)
$$
and to the asymptotic condition
$$
\lim_{r \to \infty}\varphi(r,{\hat x},\zeta;\lambda)=0.
\eqno (4.6)
$$
In Eq. (4.5), $\sigma_{g}$ is the {\it graded leading symbol} of the
boundary operator of section 2 in the local coordinates
$\left \{ {\hat x}^{k} \right \}, \left \{ \zeta_{j} \right \}$, 
and is given by
$$
\sigma_{g}(B)=\pmatrix{\Pi & 0 \cr i \Gamma^{j}\zeta_{j}
& I-\Pi \cr}.
\eqno (4.7)
$$
Roughly speaking, the above
construction uses Fourier transform and the
inward geodesic flow to obtain the ordinary differential
equation (4.4) from the Laplacian, with corresponding Fourier
transform (4.5) of the original boundary conditions.  
The asymptotic condition (4.6) picks out the solutions of Eq. (4.4)
which satisfy Eq. (4.5) with arbitrary boundary 
data $\psi'(\varphi)$ (see (2.11))
and vanish at infinite geodesic distance to the boundary. When all
the above conditions are satisfied $\forall \zeta \in 
T^{*}({\partial M}), \forall \lambda \in {\cal C}-{\bf R}_{+},
\forall (\zeta,\lambda) \not = (0,0)$ and $\forall \psi'(\varphi)
\in C^{\infty}(W',{\partial M})$, the boundary-value problem
$(P,B)$ for the Laplacian is said to be strongly elliptic with
respect to the cone ${\cal C}-{\bf R}_{+}$. 

However, when the gauge-averaging functional (3.7) is used in the
boundary condition (3.4), the work in Ref. [11] has proved that the
operator on metric perturbations takes the form of an operator of
Laplace type $P_{ab}^{\; \; \; cd}$ plus an integral operator
$G_{ab}^{\; \; \; cd}$. Explicitly, one finds [11] (with
$R_{\; bcd}^{a}$ being the Riemann curvature of the background 
geometry $(M,g)$)
$$ 
P_{ab}^{\; \; \; cd}=E_{ab}^{\; \; \; cd}(-\cstok{\ }+R)
-2 E_{ab}^{\; \; \; qf}R_{\; qpf}^{c} g^{dp}
-E_{ab}^{\; \; \; pd} R_{p}^{\; c} 
-E_{ab}^{\; \; \; cp}R_{p}^{\; d},
\eqno (4.8) 
$$
$$
G_{ab}^{\; \; \; cd}=U_{ab}^{\; \; \; cd}
+V_{ab}^{\; \; \; cd},
\eqno (4.9)
$$
where
$$
U_{ab}^{\; \; \; cd}h_{cd}(x)=-2E_{rsab}\nabla^{r} \int_{M}
T^{p}(x){\widetilde \zeta}_{\; p}^{s \; \; cd}(x,x')h_{cd}(x')dV',
\eqno (4.10)
$$
$$
h^{ab}V_{ab}^{\; \; \; cd}h_{cd}(x)=\int_{M^{2}}h^{ab}(x')
T^{q}(x){\widetilde \zeta}_{pqab}(x,x')T^{r}(x)
{\widetilde \zeta}_{\; r}^{p \; \; cd}(x,x'')h_{cd}(x'')dV'dV''.
\eqno (4.11)
$$

We now assume that the operator on metric perturbations, which is
so far an integro-differential operator defined by a kernel, is 
also pseudo-differential. This means that it can be characterized 
by suitable regularity properties obeyed by the symbol. More 
precisely, let $S^{d}$ be the set of all symbols $p(x,\xi)$ such
that [8] 
\vskip 0.3cm
\noindent
(1) $p$ is $C^{\infty}$ in $(x,\xi)$, with compact $x$ support.
\vskip 0.3cm
\noindent
(2) For all $(\alpha,\beta)$, there exist constants $C_{\alpha,\beta}$
for which
$$ \eqalignno{
\; & \left | (-i)^{\sum_{k=1}^{m}(\alpha_{k}+\beta_{k})}
{\left({\partial \over \partial x_{1}}\right)^{\alpha_{1}}
... \left({\partial \over \partial x_{m}}\right)}^{\alpha_{m}}
{\left({\partial \over \partial \xi_{1}}\right)^{\beta_{1}}
... \left({\partial \over \partial \xi_{m}}\right)}^{\beta_{m}}
p(x,\xi)\right | \cr
& \leq C_{\alpha,\beta}
{\left(1+\sqrt{g^{ab}(x)\xi_{a}\xi_{b}}\right)}^{d-\sum_{k=1}^{m}\beta_{k}},
&(4.12)\cr}
$$
for some {\it real} (not necessarily positive) value of $d$. The
associated pseudo-differential operator, defined on the Schwarz space
and taking values in the set of smooth functions on $M$ with compact
support:
$$
P: {\cal S} \rightarrow C_{c}^{\infty}(M)
$$
acts according to
$$
Pf(x) \equiv \int e^{i(x-y)\cdot \xi}p(x,\xi)f(y)\mu(y,\xi),
\eqno (4.13)
$$
where $\mu(y,\xi)$ is here meant to be the invariant integration
measure with respect to $y_{1},...,y_{m}$ and
$\xi_{1},...,\xi_{m}$. Actually, one first gives the definition 
for pseudo-differential operators $P: {\cal S} \rightarrow
C_{c}^{\infty}({\bf R}^{m})$, eventually proving that a
coordinate-free definition can be given and extended to smooth
Riemannian manifolds [8].

In the presence of pseudo-differential operators, both ellipticity
in the interior of $M$ and strong ellipticity of the 
boundary-value problem need a more involved formulation. In our
paper, inspired by the flat-space analysis in Ref. [16], we make
the following requirements.
\vskip 0.3cm
\leftline {\bf 4.1 Ellipticity in the interior}
\vskip 0.3cm
\noindent
Let $U$ be an open subset with compact closure in $M$, and
consider an open subset $U_{1}$ whose closure ${\overline U}_{1}$
is properly included into $U$: ${\overline U}_{1} \subset U$.
If $p$ is a symbol of order $d$ on $U$, it is said to be
{\it elliptic} on $U_{1}$ if there exists an open set $U_{2}$
which contains ${\overline U}_{1}$ and positive constants
$C_{0},C_{1}$ so that
$$
|p(x,\xi)|^{-1} \leq C_{1} (1+|\xi|)^{-d},
\eqno (4.14)
$$
for $|\xi| \geq C_{0}$ and $x \in U_{2}$, where $|\xi| \equiv
\sqrt{g^{ab}(x)\xi_{a}\xi_{b}}$. The corresponding operator $P$
is then elliptic.
\vskip 0.3cm
\leftline {\bf 4.2 Strong ellipticity in the absence of boundaries}
\vskip 0.3cm
\noindent
Let us assume that the symbol under consideration is
{\it polyhomogeneous}, in that it admits an asymptotic expansion
of the form
$$
p(x,\xi) \sim \sum_{l=0}^{\infty}p_{d-l}(x,\xi),
\eqno (4.15)
$$
where each term $p_{d-l}$ has the {\it homogeneity property} [16]
$$
p_{d-l}(x,t\xi)=t^{d-l}p_{d-l}(x,\xi) \; \; {\rm if} \; \;
t \geq 1 \; \; {\rm and} \; \; |\xi| \geq 1.
\eqno (4.16)
$$
The leading symbol is then, by definition,
$$
p^{0}(x,\xi) \equiv p_{d}(x,\xi).
\eqno (4.17)
$$
Strong ellipticity in the absence of boundaries is formulated in
terms of the leading symbol, and it requires that
$$
{\rm Re} \; p^{0}(x,\xi) \geq c(x) |\xi|^{d},
\eqno (4.18)
$$
where $x \in M$ and $|\xi| \geq 1$, $c$ being a positive function
on $M$. It can then be proved that the G\"{a}rding inequality holds,
according to which, for any $\varepsilon >0$,
$$
{\rm Re}(Pu,u) \geq b { \left \| u \right \| }_{{d\over 2}}^{2}
-b_{1}{ \left \| u \right \| }_{{d\over 2}-\varepsilon}^{2}
\; \; {\rm for} \; \; u \in H^{{d\over 2}}(M),
\eqno (4.19)
$$
with $b>0$, where $H^{s}(M)$ is the standard notation for Sobolev
spaces, for all $s$ [8,16].
\vskip 0.3cm
\leftline {\bf 4.3 Strong ellipticity in the presence of boundaries}
\vskip 0.3cm
\noindent
The homogeneity property (4.16) only holds for $t \geq 1$ and
$|\xi| \geq 1$. Consider now the case $l=0$, for which one obtains
the leading symbol which plays the key role in the definition 
of ellipticity. If $p^{0}(x,\xi) \equiv p_{d}(x,\xi) \equiv
\sigma_{L}(P;x,\xi)$ is not a polynomial (which corresponds
to the genuinely pseudo-differential case) while being a homogeneous
function of $\xi$, it is irregular at $\xi=0$. When $|\xi| \leq 1$,
the only control over the leading symbol is provided by
estimates of the form [16]
$$ \eqalignno{
\; & \left | (-i)^{\sum_{k=1}^{m}(\alpha_{k}+\beta_{k})}
{\left({\partial \over \partial x_{1}}\right)}^{\alpha_{1}}
... {\left({\partial \over \partial x_{m}}\right)}^{\alpha_{m}}
{\left({\partial \over \partial \xi_{1}}\right)}^{\beta_{1}}
... {\left({\partial \over \partial \xi_{m}}\right)}^{\beta_{m}}
p^{0}(x,\xi) \right| \cr
& \leq c(x) \langle \xi \rangle^{d-|\beta|}.
&(4.20)\cr}
$$
We therefore come to appreciate the problematic aspect of symbols
of pseudo-differential operators [16]. The singularity at
$\xi=0$ can be dealt with either by modifying the leading symbol
for small $\xi$ to be a $C^{\infty}$ function (at the price of
loosing the homogeneity there), or by keeping the strict 
homogeneity and dealing with the singularity at $\xi=0$ [16].

On the other hand, we are interested in a definition of strong
ellipticity of pseudo-differential boundary-value problems that
reduces to Eqs. (4.4)--(4.6) when both $P$ and the boundary 
operator reduce to the form considered in section 2. For this
purpose, and bearing in mind the occurrence of singularities
in the leading symbols of $P$ and of the boundary operator,
we make the following requirements.

Let $(P+G)$ be a pseudo-differential operator subject to boundary
conditions described by the pseudo-differential boundary
operator $\cal B$ (the consideration of $(P+G)$ rather than only
$P$ is necessary to achieve self-adjointness, as is described in
detail in Refs. [16] and [10]).
The pseudo-differential boundary-value problem
$((P+G),{\cal B})$ is strongly elliptic with respect to 
${\cal C}-{\bf R}_{+}$ if:
\vskip 0.3cm
\noindent
(I) The inequalities (4.14) and (4.18) hold;
\vskip 0.3cm
\noindent
(II) There exists a unique solution of the equation
$$
\left[\sigma_{L}\left((P+G); \left \{ {\hat x}^{k} \right \},r=0,
\left \{ \zeta_{j} \right \}, \omega \rightarrow 
-i{\partial \over \partial r} \right)-\lambda \right]
\varphi(r,{\hat x},\zeta;\lambda)=0,
\eqno (4.4')
$$
subject to the boundary conditions
$$
\sigma_{L}({\cal B})\left( \left \{ {\hat x}^{k} \right \},
\left \{ \zeta_{j} \right \} \right)\psi(\varphi)
=\psi'(\varphi)
\eqno (4.5')
$$
and to the asymptotic condition (4.6). It should be stressed that,
unlike the case of differential operators, Eq. (4.4') is not an
ordinary differential equation in general, because $(P+G)$ is
pseudo-differential.
\vskip 0.3cm
\noindent
(III) The strictly homogeneous symbols associated to $(P+G)$ and
$\cal B$ have limits for $|\zeta| \rightarrow 0$ in the respective
leading symbol norms, with the limiting symbol restricted to
the boundary which avoids the values $\lambda \not \in {\cal C}
-{\bf R}_{+}$ for all $\left \{ {\hat x} \right \}$.

Condition (III) requires a last effort for a proper understanding.
Given a pseudo-differential operator of order $d$ with leading
symbol $p^{0}(x,\xi)$, the associated strictly homogeneous symbol
is defined by [16]
$$
p^{h}(x,\xi) \equiv |\xi|^{d} p^{0} \left(x,{\xi \over |\xi|}\right)
\; \; {\rm for} \; \; \xi \not = 0.
\eqno (4.21)
$$
This extends to a continuous function vanishing at $\xi=0$ when
$d>0$. In the presence of boundaries, the boundary-value problem
$((P+G),{\cal B})$ has a strictly homogeneous symbol on the
boundary equal to (some indices are omitted for simplicity)
$$
\pmatrix{p^{h}\left( \left \{ {\hat x} \right \},r=0,
\left \{ \zeta \right \},-i{\partial \over \partial r} \right)
+g^{h}\left( \left \{ {\hat x} \right \}, 
\left \{ \zeta \right \},-i{\partial \over \partial r}
\right)- \lambda \cr
b^{h} \left( \left \{ {\hat x} \right \}, 
\left \{ \zeta \right \}, -i{\partial \over \partial r}
\right) \cr},
$$
where $p^{h},g^{h}$ and $b^{h}$ are the strictly homogeneous 
symbols of $P,G$ and $\cal B$ respectively, obtained from the
corresponding leading symbols $p^{0},g^{0}$ and $b^{0}$ via
equations analogous to (4.21), after taking into account 
the split (4.3), and upon replacing $\omega$
by $-i{\partial \over \partial r}$. The limiting symbol restricted
to the boundary (also called limiting $\lambda$-dependent boundary
symbol operator) and mentioned in condition III reads therefore [16]
$$ \eqalignno{
\; & a^{h} \left( \left \{ {\hat x} \right \}, r=0, \zeta=0,
-i{\partial \over \partial r} \right) \cr
&=\pmatrix{
p^{h} \left( \left \{ {\hat x} \right \}, r=0, \zeta=0,
-i{\partial \over \partial r} \right)
+g^{h} \left( \left \{ {\hat x} \right \}, \zeta=0,
-i{\partial \over \partial r} \right) -\lambda \cr
b^{h} \left( \left \{ {\hat x} \right \}, \zeta=0, 
-i{\partial \over \partial r} \right) \cr},
&(4.22)\cr}
$$
where the singularity at $\xi=0$ of the leading symbol in absence
of boundaries is replaced by the singularity at $\zeta=0$ of the
leading symbols of $P,G$ and $\cal B$ when a boundary occurs.  
\vskip 0.3cm
\leftline {\bf 5. Application of the strong ellipticity criterion}
\vskip 0.3cm
Let us now see how the previous conditions on the leading symbol
of $(P+G)$ and on the graded leading symbol of the boundary operator
can be used. The equation (4.4') is solved by a function $\varphi$
depending on $r, \left \{ {\hat x}^{k} \right \}, 
\left \{ \zeta_{j} \right \}$ and, parametrically, on the eigenvalues
$\lambda$. For simplicity, we write
$\varphi=\varphi(r,{\hat x},\zeta;\lambda)$, omitting indices. Since 
the leading symbol is no longer a polynomial when $(P+G)$ is 
genuinely pseudo-differential, we cannot make any further specification
on $\varphi$ at this stage, apart from requiring that it should
reduce to (here $|\zeta|^{2} \equiv \zeta_{i}\zeta^{i}$)
$$
\chi({\hat x},\zeta)e^{-r\sqrt{|\zeta|^{2}-\lambda}}
$$
when $(P+G)$ reduces to a Laplacian (and hence $\Lambda$ reduces 
to (2.3)).

The equation (4.5') involves the graded leading symbol of $\cal B$
and restrictions to the boundary of the field and its covariant
derivative along the normal direction. Such a restriction is
obtained by setting to zero the geodesic distance $r$, and hence
we write, in general form (here we denote again by $\Lambda$ the
full matrix element ${\cal B}_{21}$ in the boundary operator (3.9)),
$$
\pmatrix{\Pi & 0 \cr \sigma_{L}(\Lambda) & I-\Pi \cr}
\pmatrix{\varphi(0,{\hat x},\zeta;\lambda) \cr
\varphi'(0,{\hat x},\zeta;\lambda) \cr}
=\pmatrix{\Pi \rho(0,{\hat x},\zeta;\lambda) \cr
(I-\Pi) \rho'(0,{\hat x},\zeta;\lambda) \cr},
\eqno (5.1)
$$
where $\rho$ differs from $\varphi$, because Eq. (4.5') is
written for $\psi(\varphi)$ and $\psi'(\varphi) \not =
\psi(\varphi)$. Now Eq. (5.1) leads to 
$$
\Pi \varphi(0,{\hat x},\zeta;\lambda)=\Pi 
\rho(0,{\hat x},\zeta;\lambda),
\eqno (5.2)
$$
$$
\sigma_{L}(\Lambda)\varphi(0,{\hat x},\zeta;\lambda)
+(I-\Pi)\varphi'(0,{\hat x},\zeta;\lambda)
=(I-\Pi)\rho'(0,{\hat x},\zeta;\lambda),
\eqno (5.3)
$$
and we require that, for $\varphi$ solving Eq. (4.4') and the
asymptotic decay (4.6), with $\lambda \in {\cal C}-{\bf R}_{+}$,
Eqs. (5.2) and (5.3) can be always solved with given values of
$\rho(0,{\hat x},\zeta;\lambda)$ and $\rho'(0,{\hat x},\zeta;\lambda)$,
whenever $(\zeta,\lambda) \not = (0,0)$. The idea is now to relate,
if possible, $\varphi'(0,{\hat x},\zeta;\lambda)$ to 
$\varphi(0,{\hat x},\zeta;\lambda)$ in such a way that Eq. (5.2) can
be used to simplify Eq. (5.3). For this purpose, we consider the
function $f$ such that
$$
{\varphi'(0,{\hat x},\zeta;\lambda)\over 
\varphi(0,{\hat x},\zeta;\lambda)}=
{\rho'(0,{\hat x},\zeta;\lambda)\over 
\rho(0,{\hat x},\zeta;\lambda)}=f({\hat x},\zeta;\lambda),
\eqno (5.4)
$$
$$
\Pi({\hat x})f({\hat x},\zeta;\lambda)=f({\hat x},\zeta;\lambda)
\Pi({\hat x}).
\eqno (5.5)
$$
If both (5.4) and (5.5) hold, Eq. (5.3) reduces indeed to
$$ \eqalignno{
\; & \sigma_{L}(\Lambda)\varphi(0,{\hat x},\zeta;\lambda)
+f({\hat x},\zeta;\lambda)\Bigr(\varphi(0,{\hat x},\zeta;\lambda)
-\rho(0,{\hat x},\zeta;\lambda)\Bigr) \cr
&=f({\hat x},\zeta;\lambda)\Pi \Bigr(\varphi(0,{\hat x},\zeta;\lambda)
-\rho(0,{\hat x},\zeta;\lambda) \Bigr),
&(5.6a)\cr}
$$
and hence, by virtue of (5.2), 
$$
\Bigr[\sigma_{L}(\Lambda)+f({\hat x},\zeta;\lambda)\Bigr]
\varphi(0,{\hat x},\zeta;\lambda)=\rho'(0,{\hat x},\zeta;\lambda).
\eqno (5.6b)
$$
Thus, the strong ellipticity condition with respect to 
${\cal C}-{\bf R}_{+}$ implies in this case the invertibility of
$\Bigr[\sigma_{L}(\Lambda)+f({\hat x},\zeta;\lambda)\Bigr]$, i.e.
$$
{\rm det} \Bigr[\sigma_{L}(\Lambda)+f({\hat x},\zeta;\lambda)
\Bigr] \not = 0 \; \; \; \; \forall \lambda \in 
{\cal C}-{\bf R}_{+}.
\eqno (5.7)
$$
Moreover, by virtue of the identity
$$
\Bigr[f({\hat x},\zeta;\lambda)+\sigma_{L}(\Lambda)\Bigr]
\Bigr[f({\hat x},\zeta;\lambda)-\sigma_{L}(\Lambda)\Bigr]
=\Bigr[f^{2}({\hat x},\zeta;\lambda)-\sigma_{L}^{2}(\Lambda)\Bigr],
\eqno (5.8)
$$
the condition (5.7) is equivalent to
$$
{\rm det}\Bigr[f^{2}({\hat x},\zeta;\lambda)-\sigma_{L}^{2}(\Lambda)
\Bigr] \not = 0 \; \; \; \; \forall \lambda \in 
{\cal C}-{\bf R}_{+}.
\eqno (5.9)
$$
Since $f({\hat x},\zeta;\lambda)$ is, in general, complex-valued, one
can always express it in the form
$$
f({\hat x},\zeta;\lambda)={\rm Re}f({\hat x},\zeta;\lambda)
+i{\rm Im}f({\hat x},\zeta;\lambda),
\eqno (5.10)
$$
so that (5.9) reads eventually
$$
{\rm det}\Bigr[{\rm Re}^{2}f({\hat x},\zeta;\lambda)
-{\rm Im}^{2}f({\hat x},\zeta;\lambda)-\sigma_{L}^{2}(\Lambda)
+2i {\rm Re}f({\hat x},\zeta;\lambda)
{\rm Im}f({\hat x},\zeta;\lambda)\Bigr] \not = 0.
\eqno (5.11)
$$
In particular, when
$$
{\rm Im}f({\hat x},\zeta;\lambda)=0,
\eqno (5.12)
$$
condition (5.11) reduces to
$$
{\rm det}\Bigr[{\rm Re}^{2}f({\hat x},\zeta;\lambda)
-\sigma_{L}^{2}(\Lambda)\Bigr] \not = 0.
\eqno (5.13)
$$
A {\it sufficient condition} for strong ellipticity with respect
to the cone ${\cal C}-{\bf R}_{+}$ is therefore the negative-definiteness
of $\sigma_{L}^{2}(\Lambda)$:
$$
\sigma_{L}^{2}(\Lambda) < 0,
\eqno (5.14)
$$
so that
$$
{\rm Re}^{2}f({\hat x},\zeta;\lambda)-\sigma_{L}^{2}(\Lambda)>0,
\eqno (5.15)
$$
and hence (5.13) is fulfilled.

In the derivation of the sufficient conditions (5.11) and (5.14), the
assumption (5.5) plays a crucial role. In general, however, $\Pi$
and $f$ have a non-vanishing commutator, and hence a
$C({\hat x},\zeta;\lambda)$ exists such that
$$
\Pi({\hat x})f({\hat x},\zeta;\lambda)
-f({\hat x},\zeta;\lambda)\Pi({\hat x})
=C({\hat x},\zeta;\lambda).
\eqno (5.16)
$$
The occurrence of $C$ is a peculiar feature of the fully
pseudo-differential framework. Equation (5.3) is then equivalent to
(now we write explicitly also the independent variables in the
leading symbol of $\Lambda$)
$$ \eqalignno{
\; & \Bigr[(\sigma_{L}(\Lambda)-C)({\hat x},\zeta;\lambda)
+f({\hat x},\zeta;\lambda)\Bigr]\varphi(0,{\hat x},\zeta;\lambda) \cr
&=\rho'(0,{\hat x},\zeta;\lambda)-C({\hat x},\zeta;\lambda)
\rho(0,{\hat x},\zeta;\lambda).
&(5.17)\cr}
$$
On defining
$$
\gamma({\hat x},\zeta;\lambda) \equiv \Bigr[\sigma_{L}(\Lambda)
-C \Bigr]({\hat x},\zeta;\lambda),
\eqno (5.18)
$$
we therefore obtain strong ellipticity conditions formally analogous
to (5.7) or (5.11) or (5.13), with $\sigma_{L}(\Lambda)$ replaced
by $\gamma({\hat x},\zeta;\lambda)$ therein, i.e.
$$
{\rm det}\Bigr[\gamma({\hat x},\zeta;\lambda)
+f({\hat x},\zeta;\lambda)\Bigr] \not = 0 \; \; \forall
\lambda \in {\cal C}-{\bf R}_{+},
\eqno (5.19)
$$
which is satisfied if
$$
{\rm det}\Bigr[{\rm Re}^{2}f({\hat x},\zeta;\lambda)
-{\rm Im}^{2}f({\hat x},\zeta;\lambda)-\gamma^{2}({\hat x},\zeta;\lambda)
+2i{\rm Re}f({\hat x},\zeta;\lambda){\rm Im}f({\hat x},\zeta;\lambda)
\Bigr] \not = 0.
\eqno (5.20)
$$
\vskip 0.3cm
\leftline {\bf 6. Further applications}
\vskip 0.3cm
In the case of more mathematical interest where the operator in the 
interior of $M$ remains a Laplacian, while the boundary operator
has a pseudo-differential sector, the analysis is much simpler.
For example, two cases can be considered.
\vskip 0.3cm
\noindent
(i) If $\widetilde \Lambda$ is a pseudo-differential operator 
of order $1$, the leading symbol of the boundary operator (3.9)
can be cast in the form (cf. (4.7))
$$
\sigma_{L}({\cal B})=\pmatrix{\Pi & 0 \cr
i(T+{\widetilde T}) & I-\Pi \cr},
\eqno (6.1)
$$
where $T \equiv \Gamma^{j}\zeta_{j}$ and $\widetilde T$ results
from the occurrence of $\widetilde \Lambda$. The sufficient condition
for finding solutions of Eq. (4.5') for all $\psi'$ reads now
$$
(T+{\widetilde T})^{2}+|\zeta|^{2}I > 0 \; \; \forall
\zeta \not = 0,
\eqno (6.2)
$$
because one can simply replace $T$ with $T+{\widetilde T}$ in the
analysis of Ref. [9], if Eq. (6.1) holds. Thus, if $\widetilde \Lambda$
is chosen in such a way that
$$
(T+{\widetilde T})^{2}(\left \{ {\hat x} \right \},
\left \{ \zeta \right \}) > 0 \; \; \forall \zeta
\not = 0,
\eqno (6.3)
$$
Eq. (4.5') can always be solved with arbitrary $\psi'(\varphi)$.
The condition (6.3) can be made explicit after re-writing the
DeWitt supermetric (3.6) in the more general form
$$
E^{abcd} \equiv {1\over 2}\Bigr(g^{ac}g^{bd}+g^{ad}g^{bc}\Bigr)
+\alpha g^{ab}g^{cd}.
\eqno (6.4)
$$
Thus, on defining (with $e_{a}^{\; i}$ being a local tangent
frame on $\partial M$)
$$
\zeta_{a} \equiv e_{a}^{\; j} \; \zeta_{j},
\eqno (6.5)
$$
and introducing the nilpotent matrices
$$
(p_{1})_{ab}^{\; \; \; cd} \equiv N_{a}N_{b}\zeta^{(c} \; N^{d)},
\eqno (6.6)
$$
$$
(p_{2})_{ab}^{\; \; \; cd} \equiv N_{(a} \; \zeta_{b)} N^{c}N^{d},
\eqno (6.7)
$$
the work in Ref. [9] finds the useful formula
$$
T=-{1\over (1+\alpha)}p_{1}+p_{2},
\eqno (6.8)
$$
and this should be inserted into (6.3) to restrict the kernel of
$\widetilde \Lambda$, whose leading symbol is equal to 
$i{\widetilde T}$. The resulting restriction on $\alpha$ should 
be made compatible with the values of $\alpha$ for which the
ellipticity condition (4.14) is fulfilled in the interior of $M$.
From this point of view, one has definitely more choice than in 
the case of the local boundary operator (2.2) for an operator
of Laplace type on metric perturbations, because the values of
$\alpha$ for which the condition 
$$
T^{2}+|\zeta|^{2}I > 0 \; \; \forall \zeta 
\not = 0 
\eqno (6.9)
$$
holds (cf. (6.2)) are incompatible with the occurrence of an
operator of Laplace type on metric perturbations [9].
\vskip 0.3cm
\noindent
(ii) If $\widetilde \Lambda$ is a pseudo-differential operator of
order $d>1$ (but not necessarily integer), the leading symbol
of the boundary operator (3.9) can be expressed in the form
$$
\sigma_{L}({\cal B})=\pmatrix{\Pi & 0 \cr 
{\widehat T} & I -\Pi \cr}.
\eqno (6.10)
$$
The sufficient condition for finding solutions of Eq. (4.5')
reads instead
$$
-{\widehat T}^{2}+|\zeta|^{2}I > 0 \; \; 
\forall \zeta \not = 0.
\eqno (6.11)
$$
It is therefore sufficient to choose $\widetilde \Lambda$
in such a way that
$$
{\widehat T}^{2} < 0 \; \; \forall \zeta \not = 0.
\eqno (6.12)
$$
\vskip 0.3cm
\leftline {\bf 7. Concluding remarks}
\vskip 0.3cm
The mathematical literature and, in particular, the work by 
Grubb [14], had already considered boundary conditions of the 
form (3.8), where $(\Lambda+{\widetilde \Lambda})$ is allowed to 
be a pseudo-differential operator, but for elliptic {\it differential}
operators. In physics, however, the requirement of gauge invariance
of the boundary conditions for quantum gravity leads to an operator
on metric perturbations (see (4.8)--(4.11)) which is itself
pseudo-differential, since (3.8) is obtained from the vanishing
of the gauge-averaging functional at the boundary (see (3.4)). Our 
physical problem remains therefore original with respect to the
mathematical investigations [14,15]. Our main contributions 
are as follows.
\vskip 0.3cm
\noindent
(1) The projector condition for the boundary operator in Euclidean
quantum gravity at one loop has been derived in the form (3.15).
\vskip 0.3cm
\noindent
(2) A careful definition of strong ellipticity of pseudo-differential
boundary-value problems in Euclidean quantum gravity has been
proposed in section 4, with detailed physical applications in
section 5 (see (5.7), (5.11), (5.13), (5.19) and (5.20)), and
further mathematical examples in section 6.
\vskip 0.3cm
\noindent
In other words, we have provided a complete characterization of the
properties of the symbol of the boundary operator for which a set
of boundary conditions completely invariant under infinitesimal
diffeomorphisms are compatible with a strongly elliptic one-loop
quantum theory. The analysis of section 5 is detailed but general,
and hence has the merit (as far as we can see) of including all
pseudo-differential boundary operators for which the sufficient
conditions derived therein can be imposed.

It would be now very interesting to prove that, by virtue of the
pseudo-differential nature of $\cal B$ in (3.9), the quantum state
of the universe in one-loop semiclassical theory can be made of
surface-state type [12]. This would describe a wave function of
the universe with exponential decay away from the boundary, which
might provide a novel description of quantum physics at the
Planck length. It therefore seems
that by insisting on path-integral quantization, strong
ellipticity of the Euclidean theory and invariance principles,
new deep perspectives are in sight. These are in turn closer to what
we may hope to test, i.e. the one-loop semiclassical approximation
in quantum gravity [17]. In the seventies, such calculations could
provide a guiding principle for selecting couplings of matter fields
to gravity in a unified field theory [18]. Now they can lead instead to
a deeper understanding of the interplay 
between non-local formulations [19--21], 
elliptic theory [22,23], gauge-invariant quantization [13] 
and a quantum theory of the very early universe [17].
\vskip 0.3cm
\leftline {\bf Acknowledgment}
\vskip 0.3cm
\noindent
This work has been partially supported by PRIN97 ``Sintesi''.
Correspondence and conversations with Gerd Brubb have been very
helpful. The author is indebted to Ivan Avramidi for previous
collaboration, and to Pietro Santorelli for encouragement. 
\vskip 0.3cm
\leftline {\bf References}
\vskip 0.3cm
\noindent
\item {1.}
S. W. Hawking and G. F. R. Ellis, The Large-Scale Structure
of Space-Time, Cambridge University Press, Cambridge, 1973.
\item {2.}
E. Witten, Not. Amer. Math. Soc. 45 (1998) 1124.
\item {3.}
B. S. DeWitt, Dynamical Theory of Groups and Fields,
Gordon and Breach, New York, 1965.
\item {4.}
B. S. DeWitt, 
in Relativity, Groups and Topology II, edited by B. S. DeWitt
and R. Stora, North-Holland, Amsterdam, 1984. 
\item {5.}
R. P. Feynman, Rev. Mod. Phys. 20 (1948) 367. 
\item {6.}
C. W. Misner, Rev. Mod. Phys. 29 (1957) 497. 
\item {7.}
S. W. Hawking, 
in General Relativity, an Einstein Centenary Survey, edited by 
S. W. Hawking and W. Israel, Cambridge University Press, 
Cambridge, 1979. 
\item {8.}
P. B. Gilkey, Invariance Theory, the Heat Equation and the
Atiyah--Singer Index theorem, Chemical Rubber Company,
Boca Raton, 1995.
\item {9.}
I. G. Avramidi and G. Esposito, Commun. Math. Phys.
200 (1999) 495. 
\item {10.}
G. Esposito, Class. Quantum Grav. 16 (1999) 1113. 
\item {11.}
G. Esposito, Class. Quantum Grav. 16 (1999) 3999. 
\item {12.}
M. Schr\"{o}der, Rep. Math. Phys. 27 (1989) 259. 
\item {13.}
G. Esposito and C. Stornaiolo, Int. J. Mod. Phys. A15
(2000) 449.
\item {14.}
G. Grubb, Ann. Scuola Normale Superiore Pisa, Ser. IV,
1 (1974) 1.
\item {15.}
P. B. Gilkey and L. Smith, J. Diff. Geom. 18 (1983) 393. 
\item {16.}
G. Grubb, Functional Calculus of Pseudodifferential
Boundary Problems, Birkh\"{a}user, Boston, 1996.
\item {17.}
G. Esposito, A. Yu. Kamenshchik and G. Pollifrone, Euclidean
Quantum Gravity on Manifolds with Boundary, Fundamental
Theories of Physics, Vol. 85, Kluwer, Dordrecht, 1997.
\item {18.}
G. 't Hooft and M. Veltman, Ann. Inst. H. Poincar\'e 20
(1974) 69.
\item {19.}
J. W. Moffat, Phys. Rev. D41 (1990) 1177. 
\item {20.}
D. Evens, J. W. Moffat, G. Kleppe and R. P. Woodard,
Phys. Rev. D43 (1991) 499. 
\item {21.}
V. N. Marachevsky and D. V. Vassilevich, Class. Quantum Grav.
13 (1996) 645. 
\item {22.}
I. G. Avramidi and G. Esposito, Class. Quantum Grav. 15 (1998) 281.
\item {23.}
J. S. Dowker and K. Kirsten, Class. Quantum Grav. 16 (1999) 1917.

\bye